# IMPACTO REDISTRIBUTIVO DA REFORMA DA TRIBUTAÇÃO DO CONSUMO NO BRASIL: SIMULAÇÕES BASEADAS NO PLP 68/2024

*Rozane Bezerra de Siqueira\**
*José Ricardo Bezerra Nogueira\**
*Carlos Feitosa Luna\*\**[1]

*(18 de novembro de 2024)*

## 1. Introdução

O objetivo deste estudo é estimar as implicações para a alíquota de referência e para a distribuição da carga tributária entre as famílias da reforma da tributação do consumo conforme delineada na versão do projeto de lei complementar (PLP) 68 aprovada pela Câmara dos Deputados em julho de 2024 (PLP 68, 2024). Trata-se de uma atualização de parte das simulações apresentadas em Siqueira, Nogueira e Luna (2024), realizadas meses antes da aprovação do projeto de regulamentação.

A metodologia utilizada e as fontes de dados são as mesmas do estudo anterior. Apenas os parâmetros relativos às alíquotas tributárias e suas bases de incidência foram modificados para melhor refletir o projeto aprovado na Câmara. Assim, para melhor avaliar e compreender os resultados aqui apresentados, recomenda-se ao leitor interessado consultar Siqueira, Nogueira e Luna (2024).

Os resultados indicam que a base exclusiva da alíquota de referência do novo tributo criando pela reforma representa apenas cerca de 46% do consumo das famílias. O restante do consumo estará sujeito à isenção, à alíquota reduzida, à regime específico de tributação ou ao Imposto Seletivo. Apesar disso, o efeito equalizador da reforma é modesto e não muito distante de um sistema com alíquota uniforme e pagamento de cashback para os mais pobres. Por outro lado, o impacto das isenções e alíquotas favorecidas sobre a alíquota de referência é drástico, podendo esta ultrapassar em muito o patamar de 30%.

## 2. Reforma simulada com base no PLP 68/2024

O PLP 68/2024 regulamenta a Contribuição sobre Bens e Serviços (CBS), da União; o Imposto sobre Bens e Serviços (IBS), dos estados e municípios; e o Imposto Seletivo (IS), que incidirá sobre produtos considerados nocivos à saúde e ao meio ambiente.

Quanto à estrutura de alíquotas, o texto prevê alíquota zero para alguns bens e serviços, uma alíquota de referência, uma alíquota de 40% da alíquota de referência e outra de 70% da alíquota de referência, além das alíquotas dos regimes específicos de tributação e do IS. É também estabelecido a devolução de parte dos tributos – o chamado *cashback* – para as famílias inscritas no Cadastro Único.

---

[1] *Departamento de Economia, Universidade Federal de Pernambuco; **Fundação Oswaldo Cruz/FIOCRUZ. Endereço para correspondência: rozane_siqueira@yahoo.com.br.

Nas simulações, procuramos reproduzir o mais próximo possível – na medida em que os dados permitem – a reforma delineada no texto do PLP 68/2024 aprovado na Câmara. A seguir descrevemos a reforma simulada, começando pela definição das alíquotas – e suas respectivas bases de incidência – que são tomadas como parâmetros no cálculo do cashback e da alíquota de referência.

Alíquota zero

É aplicada sobre a **cesta básica de alimentos** especificada no Anexo I do texto do PLP 68/2024; **produtos hortícolas, frutas e ovos** relacionados no Anexo XV; **medicamentos** relacionados no Anexo XIV; **produtos de cuidados básicos à saúde menstrual**; **plantas**; **livros**; e **transporte público coletivo de passageiros rodoviário, metroviário, e ferroviário de caráter urbano, semiurbano e metropolitano**.[2]

Alíquotas dos regimes específicos

Os regimes específicos de tributação são diferenciados em relação ao modelo de cobrança da CBS e do IBS, que são do tipo Imposto sobre Valor Agregado (IVA), porém suas alíquotas não são necessariamente favorecidas relativamente à alíquota de referência. No geral, o PLP 68/2024 prevê a manutenção da carga tributária atual para os itens dos regimes específicos.

Assim, para os regimes específicos que simulamos, utilizamos as alíquotas tributárias efetivas sobre as famílias estimadas para o sistema vigente de tributos indiretos. A metodologia de cálculo dessas alíquotas é descrita detalhadamente em Siqueira, Nogueira e Luna (2024). Cabe ressaltar que as estimativas partem das receitas efetivamente arrecadadas pelo governo de cada setor de atividade (líquidas de subsídios) e leva em conta os efeitos da tributação de insumos.

As seguintes alíquotas foram atribuídas: **gasolina**: 33%, no cálculo com o imposto por dentro da base (equivalente a 49,3% no cálculo por fora); **outros produtos do refino do petróleo e etanol**: 27% por dentro (37% por fora); **serviços financeiros**: 18% por dentro (22% por fora); **bares e restaurantes**: 14% por dentro (16,3% por fora); **serviços de hotelaria, parques de diversão e agências de turismo**: 14% por dentro (16,3% por fora); **transporte público coletivo de passageiros rodoviário, ferroviário, hidroviário intermunicipal e interestadual**: 20% por dentro (25% por fora).

---

[2] Vale ressalvar que no PLP 68/2024 essa categoria de transporte se encontra entre os regimes especiais de tributação, porém, com alíquota zero.

Alíquotas do Imposto Seletivo

O PLP 68/2024 não fixa as alíquotas do imposto seletivo, que serão definidas em nova lei complementar. De forma que nas simulações essas alíquotas tiveram que ser arbitradas. No caso de **bebidas alcóolicas**, fixamos a alíquota do IS em 19% (por fora), que em conjunto com o IVA reproduz aproximadamente a alíquota efetiva do sistema vigente. A mesma alíquota de 19% foi atribuída para **produtos fumígenos**. Para **veículos e embarcações** a alíquota utilizada nas simulações foi de 5% e para **bebidas açucaradas** foi fixada em 3%. Por fim, a alíquota de 15% foi arbitrada para **apostas e loterias.** Cabe ressaltar que o IS faz parte da base do IVA, o que significa que a alíquota efetiva de um item sujeito ao IS é maior do que a soma das alíquotas legais do IVA e o IS.

Alíquota de 40% da alíquota de referência

Aplicada a: **alimentos relacionados no Anexo VII** do texto do PLP 68/2024; **serviços de educação**; **serviços de saúde**; **medicamentos**; **sabões de toucador, dentifrícios, escovas de dente, papel higiênico, água sanitária e sabões em barra**; **dispositivos médicos**; **dispositivos de acessibilidade para pessoas com deficiência**; **produções artísticas, culturais e atividades desportivas**; **planos de assistência à saúde**; e **aluguel de bem imóvel**.

Cabe a ressalva de que plano de assistência à saúde e aluguel de imóvel estão sujeitos a regime específico de tributação, porém, a alíquota proposta no PLP 68/2024 é 40% da alíquota de referência. No caso do aluguel de imóvel é previsto ainda um 'redutor social' de R$ 400,00 (em reais de hoje) da base de cálculo do imposto, que foi levado em conta em nossas simulações.

Alíquota de 70% da alíquota de referência

Aplicada sobre: **serviços de profissão intelectual, de natureza científica, literária ou artística;** e **planos de saúde para animal doméstico.**

Alíquota de referência e *cashback*

De acordo com a reforma aprovada, a alíquota de referência deve ser fixada de forma que o novo sistema reproduza a carga tributária dos tributos que serão extintos. Portanto, o cálculo da alíquota de referência requer uma estimativa da carga tributária pré-reforma. Usando a metodologia de Siqueira, Nogueira e Luna (2024), estimamos que os tributos indiretos representam 20,1% da despesa (monetária) de consumo das famílias.[3]

---

[3] Neste estudo, optamos por incluir a categoria 'serviços domésticos' na variável despesa de consumo usada no cálculo da carga tributária. Em Siqueira, Nogueira e Luna (2024) essa categoria foi excluída por não fazer parte da base do IVA. Isso explica o fato da carga tributária agregada apresentada aqui aparecer levemente menor do que no estudo anterior.

A alíquota de referência depende também do montante de *cashback* pago às famílias, o qual, por sua vez, é uma função da alíquota de referência, dadas as demais alíquotas. O *cashback* previsto no PLP 68/2024 é para famílias com renda familiar per capita de até meio salário mínimo. Os percentuais de devolução do imposto são de 100% da CBS e 20% do IBS no caso de botijão de gás, energia elétrica, gás natural, água e esgoto, e 20% da CBS e do IBS nos demais casos, exceto produtos sujeitos ao IS.

Nas simulações, o percentual de 46,6% foi usado para calcular o cashback relativo à tributação de gás, energia elétrica, água e esgoto, o qual foi estimado considerando as participações relativas da CBS e do IBS na alíquota do novo tributo.

Para contornar o problema da dependência mútua entre a alíquota de referência e o *cashback*, primeiro calculamos o *cashback* utilizando uma estimativa da alíquota de referência na ausência de *cashback*, e então usamos o *cashback* estimado para recalcular a alíquota de referência. Esse processo foi repetido até que as alterações na alíquota de referência deixaram de ser significativas.

Assim, obtivemos que a alíquota de referência associada ao desenho de reforma acima descrito é 27,5% no cálculo com o imposto por dentro, equivalente a 37,9% no cálculo por fora. Cabe lembrar que esta alíquota é neutra em termos de arrecadação.

Como mencionado acima, a carga tributária resultante do sistema simulado é de aproximadamente 20% do consumo das famílias, que é a mesma carga gerada pela aplicação de um IVA com alíquota uniforme de 25% (calculada por fora).

Deve-se ter em mente também que no sistema vigente de tributos indiretos itens importantes no consumo das famílias e que são insumos produtivos básicos estão sujeitos a alíquotas efetivas mais elevadas do que a alíquota de referência aqui estimada, a saber: energia elétrica, com alíquota efetiva de 51% por fora; gasolina, 49%; e telecomunicação, 42% (ver Siqueira, Nogueira e Luna, 2024).

## 3. Consequências da seletividade para a alíquota de referência

As tabelas 1 e 2 apresentam estimativas das consequências das isenções e alíquotas diferenciadas para a base e a magnitude da alíquota de referência.

Para construir a Tabela 1 agrupamos os bens e serviços consumidos pelas famílias de acordo com o tratamento tributário recebido. A tabela mostra a participação média de cada grupo na despesa domiciliar, por quinto de despesa total.[4] Olhando para a média global (última coluna), verifica-se que as despesas sobre as quais incide a alíquota de referência de forma exclusiva – ou seja, sem a incidência conjunta do IS – representam menos da metade da despesa total das famílias: 45,7%. Por sua vez, a base do IS representa 9,7% da despesa total. Cabe observar que o grupo tributado à alíquota de referência tem um peso maior no orçamento dos mais pobres do que dos mais ricos. O contrário acontece com as despesas sujeitas ao IS.

---

[4] Os quintos são construídos ordenando-se os domicílios de acordo com a despesa total de consumo domiciliar *per capita* (incluindo a despesa não-monetária).

No caso das despesas com alíquota favorecida, observa-se que o peso orçamentário da cesta básica cai fortemente à medida que a despesa total aumenta. Comportamento semelhante ocorre com a categoria 'outras despesas com alíquota zero', que inclui transporte público urbano e medicamentos. Por outro lado, para os demais grupos de despesa com alíquota reduzida a participação no orçamento das famílias aumenta com a despesa total. Em particular, destaca-se o grupo com alíquota reduzida em 60% da alíquota de referência, com peso duas vezes maior na cesta de consumo dos 20% mais ricos do que dos 20% mais pobres.

**Tabela 1**: Participação das despesas agrupadas por tipo de tratamento tributário no orçamento das famílias (%)

|  | Quintos de despesa total | | | | | |
|---|---|---|---|---|---|---|
|  | 1 | 2 | 3 | 4 | 5 | Total |
| Cesta básica | 15,0 | 12,5 | 10,6 | 8,6 | 4,3 | 7,5 |
| Outras com alíquota zero | 7,4 | 7,4 | 6,8 | 5,8 | 3,8 | 5,2 |
| Alíquota de referência | 52,7 | 50,4 | 48,2 | 47,0 | 42,6 | 45,7 |
| 40% da alíquota de referência | 7,9 | 9,6 | 11,4 | 13,1 | 16,3 | 13,8 |
| 70% da alíquota de referência | 0,1 | 0,3 | 0,5 | 0,7 | 1,2 | 0,8 |
| Aluguel de imóvel | 2,9 | 4,2 | 4,4 | 3,9 | 3,3 | 3,7 |
| Regime específico | 10,6 | 11,1 | 12,1 | 12,7 | 15,2 | 13,6 |
| Imposto Seletivo | 3,3 | 4,5 | 5,9 | 8,2 | 13,3 | 9,7 |
| Total | 100,0 | 100,0 | 100,0 | 100,0 | 100,0 | 100,0 |

Fonte: Cálculo dos autores a partir da POF 2017-2018.

A Tabela 2 mostra a variação marginal na alíquota de referência resultante da eliminação do tratamento diferenciado especificado. Observa-se que a isenção da cesta básica e a redução de 60% da alíquota de referência têm impactos semelhantes sobre a alíquota de referência, cerca de 5,0 pontos percentuais.[5] Por sua vez, a manutenção da carga tributária vigente sobre 'bares e restaurantes', 'serviços de hotelaria e parques de diversão´ e 'agências de turismo' resulta em um aumento de 2,0 p.p. na alíquota de referência. O IS, por outro lado, conforme aqui simulado, permite reduzir a alíquota de referência em 2,5 p.p. Por fim, a tabela mostra que o pagamento de *cashback* tem um impacto de 1,5 p.p. sobre a alíquota de referência.

---

[5] Vale mencionar que carnes e peixes representam aproximadamente 45% da despesa com a cesta básica de alimentos.

**Tabela 2:** Impacto das alíquotas favorecidas, do Imposto Seletivo, e do *cashback* sobre a alíquota de referência

| | | Variação em p.p. |
|---|---|---|
| Alíquota de referência sem cashback | 36,4 | |
| Sem isenção da cesta básica de aimentos | 31,4 | -5,0 |
| Sem isenção de medicamentos e transporte público urbano | 34,4 | -2,0 |
| Sem alíquota reduzida de 40% | 30,9 | -5,5 |
| Sem alíquota reduzida de 70% | 36,2 | -0,2 |
| Sem regime específico bares e restaurantes, hotelaria e turismo | 34,4 | -2,0 |
| Sem Imposto Seletivo | 38,9 | 2,5 |
| Alíquota de referência com cashback | 37,9 | 1,5 |

Fonte: Cálculo dos autores a partir da POF 2017-2018.

## 4. Efeitos redistributivo da reforma

Esta seção examina o impacto redistributivo da reforma acima desenhada com base no PLP 68/2024, tomando como referência o sistema vigente de tributos indiretos.

Com o objetivo de destacar a contribuição das alíquotas favorecidas para o efeito equalizador da reforma, simulamos dois outros cenários: um IVA uniforme sobre todos os bens e serviços (sem imposto seletivo nem cashback) e uma versão do PLP 68/2024 em que a cesta básica de alimentos é tributada à alíquota de referência e a receita extra é utilizada para financiar uma transferência de montante fixo para todos os indivíduos na população. Nos três cenários a carga tributária global sobre consumo das famílias, líquida de *cashback* ou transferência, é a mesma do sistema corrente.

A Tabela 3 apresenta, para o sistema vigente e para as reformas simuladas, o montante médio, por domicílio, de tributos pagos sobre o consumo, por quinto de despesa total. A tabela também mostra a variação na tributação de cada quinto em relação ao sistema corrente, e essa variação como fração na despesa total de consumo.

**Tabela 3**: Impacto redistributivo da reforma e sistemas alternativos

| | Domicílios ordenados pela despesa total per capita | | | | |
|---|---|---|---|---|---|
| | Quintos | | | | |
| | 1 | 2 | 3 | 4 | 5 |
| **Sistema vigente** | | | | | |
| Tributação (R$/mês) | 197 | 330 | 457 | 652 | 1.389 |
| Depesa (monetária) total (R$/mês) | 759 | 1.417 | 2.114 | 3.200 | 7.543 |
| **IVA uniforme** | | | | | |
| Tributação (R$/mês) | 152 | 284 | 425 | 643 | 1.515 |
| Variação na tributação (R$/mês) | -45 | -46 | -32 | -9 | 126 |
| Variação na tributação / despesa total | -0,059 | -0,032 | -0,015 | -0,003 | 0,017 |
| **PLP 68/2024** | | | | | |
| Tributação (R$/mês) | 128 | 259 | 407 | 647 | 1.573 |
| Variação na tributação (R$/mês) | -69 | -71 | -50 | -5 | 184 |
| Variação na tributação / despesa total | -0,091 | -0,050 | -0,023 | -0,002 | 0,024 |
| **PLP 68 s/ isenção da cesta c/ transferência universal** | | | | | |
| Tributação (R$/mês) | 89 | 243 | 408 | 666 | 1609 |
| Variação na tributação (R$/mês) | -108 | -87 | -49 | 14 | 220 |
| Variação na tributação / despesa total | -0,142 | -0,061 | -0,023 | 0,004 | 0,029 |

Nota: Os valores monetários estão expressos em reais de 2018.
Fonte: Cálculo dos autores a partir da POF 2017-2018.

Como já observado em Siqueira, Nogueira e Luna (2024), a substituição da estrutura corrente de tributos indiretos por um IVA com alíquota única tem um efeito levemente progressivo, reduzindo a tributação dos domicílios de renda mais baixa. Isso significa que a diferenciação das alíquotas no sistema vigente favorece as classes de renda mais alta, o que, em grande medida, se deve à subtributação de serviços.

A Reforma baseada no PLP 68/2024, apesar das isenções, múltiplas alíquotas e *cashback*, não tem um efeito equalizador muito mais significativo, como se poderia esperar, gerando um ganho médio para os 20% mais pobres equivalente a 9% da despesa domiciliar, comparado com um ganho de 6% no caso do IVA uniforme e sem cashback. Nos dois casos, apenas os 20% mais ricos sofrem aumento de tributação, em montante equivalente a cerca de 2% da despesa domiciliar.

Por fim, pode-se observar que a simulação que substitui a isenção da cesta básica, no contexto do PLP 68/2024, por uma transferência universal de montante fixo é mais progressiva do que o modelo simulado do PLP 68/2024, produzindo um ganho para os 20% mais pobres equivalente a 14% da despesa domiciliar total, em média. Como observa Siqueira, Nogueira e Luna (2024), esse resultado evidencia o quanto o gasto tributário envolvido na isenção da cesta básica é mal focalizado.

## 5. Conclusão

Este estudo atualiza as simulações apresentadas em Siqueira, Nogueira e Luna (2024) dos efeitos redistributivos e sobre a alíquota de referência da reforma da tributação do consumo. Esta nova versão do estudo é baseada no texto do PLP 68 aprovado na Câmara dos Deputados em julho de 2024. A metodologia e dados utilizados são os mesmos do estudo anterior.

Em termos qualitativos, as conclusões não são alteradas em relação ao primeiro trabalho. As simulações indicam que apesar das isenções e outros tratamentos tributários diferenciados, além do *cashback*, a reforma tributária tem um efeito equalizador modesto sobre a distribuição da carga tributária entre as famílias. No entanto, o impacto das isenções e alíquotas reduzidas sobre a base de incidência e a magnitude da alíquota de referência é considerável. Os bens e serviços tributados exclusivamente por essa alíquota representam apenas 45,7% do consumo das famílias e o nível estimado da alíquota neutra é 37,9%.

Convém ressalvar que devido às limitações dos dados e, ainda, a indefinição de alguns parâmetros da reforma tributária em tramitação no Congresso Nacional – como as alíquotas do IS e as alíquotas que serão efetivamente aplicadas ao itens dos regimes específicos – a alíquota de referência estimada neste estudo deve ser vista como oferecendo apenas uma ideia de ordem de magnitude.

Vários elementos que ainda serão definidos, ou que dependem de dinâmicas criadas no processo de implementação da reforma, podem influenciar o nível da alíquota de referência. Alguns fatores podem levar à redução dessa alíquota, tais como calibração adequada das alíquotas do IS ou de itens dos regimes específicos, e alguma cumulatividade remanescente (associada à tributação de combustíveis pelo IS, por exemplo).

Por outro lado, há circunstâncias não consideradas na simulação que podem levar a aumento da alíquota de referência, como a presença de empresas tributadas pelo Simples e atividades de sonegação e elisão fiscal. Nesse sentido, vale notar que a reforma reduz significativamente a tributação de energia elétrica e telecomunicação, com a receita tendo que ser recuperada de setores em que a arrecadação é administrativamente mais difícil e a demanda mais elástica.

**Referências**